\documentclass[preprint,showpacs,preprintnumbers,amsmath,amssymb]{revtex4}
\usepackage{bm}
\begin{document}

\title[Geodesic deviation in $pp$-wave spacetimes]{Geodesic deviation in $pp$-wave spacetimes of quadratic curvature gravity}

\author{Edgard C. de Rey Neto}

\email{{edgard@das.inpe.br}}
\affiliation{Instituto Nacional de Pesquisas Espaciais - Divis\~ao de
Astrof\'\i sica \\ Av. dos Astronautas 1758, S\~ao Jos\'e dos
Campos, 12227-010 SP, Brazil}

\date{September 05, 2003}

\begin{abstract}
We write the equation of geodesic deviations in the spacetime of $pp$-waves in terms of the Newman-Penrose scalars and apply it to study gravitational waves in quadratic curvature gravity. We show that quadratic curvature gravity $pp$-waves can have a transverse helicity-0 polarization mode and two transverse helicity-2 general relativity-like wave polarizations. A concrete example is given in which we analyze the wave polarizations of an exact impulsive gravitational wave solution to quadratic curvature gravity.

\end{abstract}


\maketitle

\newpage

\section{Introduction}

In the present days, we assist a increasing worldwide activity to detect gravitational radiation. These efforts will certainly lead to a birth of the gravitational wave astronomy in the beginning of the present century.  
Consequently, there is a growing theoretical interest in the description of gravitational waves emission and detection. Besides the information on astrophysical process  that can be gained from gravitational wave observations, the direct detection of gravitational waves will be also of great importance to the understanding of the nature of gravity. Within this context, although the common belief that the linearized theory is sufficient to describe gravitational waves, exact solutions of gravity equations which represents gravitational waves also deserves to be investigated. One reason to this is that  the linearization of gravity equations can hide important features contained in the exact nonlinear systems which could be useful to the understanding the global content of the gravitational theories involved. For instance, exact radiative solutions to nonlinear curvature gravities are easily obtained for $pp$-wave metrics while, the contribution of the nonlinear curvature terms to linearized quadratic gravity waves are obscured by the linear approximation~\cite{RAA1,RAA2}.
Moreover, as pointed out in~\cite{Chris1}, the nonlinear content of Einstein's gravity cannot be neglected in many physically interesting cases such as the  gravitational emission in the coalescence of binary systems. Also, in~\cite{CVV}, exact solutions to general relativity which represents spin-1 gravitational waves are found.

In the present work we study the geodesic deviations in $pp$-wave spacetimes in the framework of quadratic curvature gravity. The plane fronted gravitational waves with parallel rays, $pp$-waves, are spacetimes which admits a covariantly constant null vector field. These spacetimes represents plane gravitational waves which propagates with the fundamental velocity $c$. They constitute a subclass of the general Kundt class of exact plane gravitational waves~\cite{KSMH}.
Here, we are interested in obtaining the effect of quadratic gravity waves on geodetic test particles. Then, we study the geodesic deviations in $pp$-wave spacetimes which we know to be simple exact solutions to quadratic curvature gravity~\cite{buchdahl}. The existence of exact $pp$-wave solutions to quadratic gravity leads to the question of the existence of solutions for the general Kundt class of exact plane waves and also for the Robinson-Trautman class of exact spherical gravitational waves in quadratic gravity. However, the investigations of this subject are beyond the scope of the present work. The choice of the $pp$-wave spacetime metric also excludes the non-null wave like solutions of quadratic gravity, such as those found in~\cite{MR1}.  
 
The equation of geodesic deviations gives the relative accelerations between free test particles falling in a gravitational field and is a cornerstone to the understanding of the physical effects of the gravitation~\cite{P1}, being the basis of almost all prospects for the gravitational wave detection. Geodesics and geodesic deviations in spacetimes of impulsive $pp$-waves of general relativity are rigorously studied in~\cite{Stein} by using the concepts of the Colombeau algebras to handle the nonlinear products of distributions. We do not follow here the same approach developed in~\cite{Stein}, rather than, we write the equation of geodesic deviations in a orthonormal tetrad basis by projecting the components of the Riemann tensor on this basis. Then, we obtain the relative accelerations of nearby test particles as function of the Newman-Penrose (NP) quantities. Another serie of recent papers which deals with geodesic deviations in spacetimes of general relativity and, in particular with the field of $pp$-waves deserves to be cited here~\cite{kerner}. However, we stress that our approach to the issue of geodesic deviations in the field of $pp$-waves also differ from that carried out in the above references.

The plane of the paper is the following. In the section~\ref{secnp} we write a NP null tetrad in the spacetime of generic $pp$-waves and the nonvanishing NP quantities. In the section~\ref{secgd} we write the geodesic deviation equation in terms of the NP quantities. Then, we reduce to the case of $pp$-waves and obtain the wave polarizations which can arise for these metrics. In the section~\ref{secqg}, by considering $pp$-waves which are solution of non-linear Lagrangian field equations, we obtain the geodesic deviations as such as, the wave polarization that can be found in quadratic curvature gravity. We also analyze a concrete example given by the gravitational wave solution to quadratic curvature gravity obtained in~\cite{RAA1}.

\section{Newman-Penrose formalism for generic $pp$-waves}
\label{secnp}
In this section we extend the method employed by Hayashi and Samura in~\cite{HaySam2} to construct a null tetrad frame for cylindrically symmetric gravitational shock waves, to the case of waves with arbitrary amplitude and arbitrary dependence on angular coordinates. Then, we write the non-necessarily  null NP quantities for a generic $pp$-wave.  

We start with a $pp$-wave described by the line element spacetime~$(\hbar=c=1)$:
\begin{equation}
ds^2=-dudv+H(x^k,u)du^2+g_{kl}dx^kdx^l,
\label{lin1}
\end{equation}
where $u=t-z$, $v=t+z$ and $x^k$ are generalized coordinates in the $2$-dimensional space perpendicular to the propagation direction, the so called transverse space. The Latin indices run over the transverse space dimensions $k=1,2$. The line element~(\ref{lin1}) clearly represents a plane wave propagating in the $z$ direction with the fundamental velocity. Let us write $H=H(x_\perp,u)$ where $x_\perp$ is a point in the transverse space. The particular form of $H(x_\perp,u)$ will depend on the source term as such as on the underling theoretical model. 

The non-necessarily  null Christoffel symbols for the spacetime defined by~(\ref{lin1}) are 

\begin{eqnarray}
{\Gamma}^k_{lm}={\hat{\Gamma}^k}_{lm}\quad,\quad{\Gamma}^v_{ul}={\Gamma}^v_{lu}=-\frac{\partial H}{\partial x^l}(x_\perp,u), \nonumber \\ 
{\Gamma}^v_{uu}=-\frac{\partial H}{\partial u}(x_\perp,u)\quad,\quad{\Gamma}^k_{uu}=-\frac{1}{2}g^{kl}\frac{\partial H}{\partial x^l}(x_\perp,u), 
\label{Gamma}
\end{eqnarray}
where ${\hat{\Gamma}^k}_{lm}$ are the transverse space Christoffel symbols. In the spacetime~(\ref{lin1}) we have
\begin{equation}
R=0\quad;\quad R_{\mu\nu}R^{\mu\nu}=0\quad {\rm and}\quad R_{\mu\nu\alpha\beta}R^{\mu\nu\alpha\beta}=0,
\label{0inv2} 
\end{equation}
and the nonvanishing components of the Ricci tensor are
\begin{equation}
R_{kl}=\hat{R}_{kl}\quad {\rm and}\quad R_{uu}=-\frac{1}{2}\nabla_{\perp}^2H(x_\perp,u),
\label{Rab}
\end{equation}
where $\nabla_\perp^2$ is the Laplacian in the transverse space. 
In flat transverse space we have $\hat{R}_{kl}=0$. The geodesic equations are
\begin{subequations}
\label{geoeq}
\begin{equation}
\ddot{u}=0,
\end{equation}
\begin{equation}
\ddot{v}=2\frac{\partial H}{\partial x^k}\dot{x}^k\dot{u}+\frac{\partial H}{\partial u}\dot{u}^2,
\end{equation}
\begin{equation}
\ddot{x}^k=\frac{1}{2}g^{kl}\frac{\partial H}{\partial x^l}\dot{u}^2-{\hat{\Gamma}^k}_{lm}\dot{x}^l\dot{x}^m,
\end{equation}
\end{subequations}
where the dot means derivative with respect to the affine parameter. 

Now, we consider flat transverse space and define cylindrical coordinates on it by writing  $x^1=\rho\cos(\phi)$, $x^2=\rho\sin(\phi)$. This choice makes apparent the radial and angular aspects of the $pp$-wave metric. The line element~(\ref{lin1}) takes the form
\begin{equation}
ds^2=-dudv+H(\rho,\phi,u)du^2+d\rho^2+\rho^2d\phi^2.
\label{lin2}
\end{equation}
The geodesic equations becomes
\begin{subequations}
\label{geo2}
\begin{equation}
\ddot{u}=0, 
\end{equation}
\begin{equation}
\ddot{v}=\frac{\partial H}{\partial u}\dot{u}^2+2\frac{\partial H}{\partial\rho}\dot{\rho}\dot{u}+2\frac{\partial H}{\partial\phi}\dot{\phi}\dot{u}, 
\end{equation}
\begin{equation}
\ddot{\phi}=\frac{1}{2\rho^2}\frac{\partial H}{\partial\phi}\dot{u}^2-\frac{2}{\rho}\dot{\phi}\dot{\rho},
\end{equation}
\begin{equation}
\ddot{\rho}=\frac{1}{2}\frac{\partial H}{\partial\rho}\dot{u}^2+\rho\dot{\phi}^2.
\end{equation}
\end{subequations}

Let us write a point in the spacetime with coordinates $(u,v,\rho,\phi)$ as $\bm{x}$.
To construct a null tetrad basis we need to know the first derivative of the $\bm{x}$ with respect to the affine parameter. This is obtained by integrating the geodesic equations once with respect to the affine parameter and holding the transverse coordinates fixed. 

The first of equations~(\ref{geo2}) implies that $u=a_0s+b_0$. Without loss of generality, we choose $b_0=0$ and $a_0=1$ so that $u$ can be taken as the affine parameter itself. Thus,

\begin{subequations}
\label{lrho}
\begin{equation}
\dot{u}=1,
\end{equation}
\begin{equation}
\dot{v}=2H-{\int_{u_0}^u\frac{\partial H}{\partial u}du},
\end{equation}
\begin{equation}
\dot{\phi}=\frac{1}{2\rho^2}{\int_{u_0}^u\frac{\partial H}{\partial \phi}du},
\end{equation}
\begin{equation}
\dot{\rho}=\frac{1}{2}{\int_{u_0}^u\frac{\partial H}{\partial\rho}du}+\frac{1}{4\rho^3}\left({\int_{u_0}^u\frac{\partial H}{\partial\phi}du}\right)^2,
\end{equation}
\end{subequations}
where we neglect the inessential constants.

Now, we define the vector $\bm{l}$ by the vector whose contravariant components are $l^\mu~\equiv~(\dot{u},\dot{v},\dot{\rho},\dot{\phi})$, $\mu=0,1,2,3$. By noting that 
\begin{equation}
\dot{\rho}^2+\rho^2\dot{\phi}^2=H-{\int_{u0}^u\frac{\partial H}{\partial u}du},
\label{pcond}
\end{equation}
(take the total derivative with respect to $u$ of both sides of~(\ref{pcond}))
one can show that $l_\mu l^\mu=0$. 

The equations~(\ref{lrho}) defines a null tangent vector to the spacetime~(\ref{lin2}).
We can build a Newman-Penrose tetrad by taking the vector $\bm{l}$ above, a null real vector $\bm{k}$ orthogonal to $\bm{l}$
and two null complex conjugated vectors $\bm{m}$ and $\bm{\overline{m}}$ required to satisfy the orthogonality conditions:
\begin{equation}
\label{orthc}
\bm{l}\cdot\bm{m}=\bm{l}
\cdot\bm{\overline{m}}=\bm{k}\cdot\bm{m}=\bm{k}\cdot\bm{\overline{m}}=0,
\end{equation}
the null conditions:
\begin{equation}
\label{nullc}
\bm{l}\cdot\bm{l}=\bm{k}\cdot\bm{k}=\bm{m}\cdot\bm{m}=\bm{\overline{m}}\cdot\bm{\overline{m}}=0 
\end{equation}
and the normalization conditions:
\begin{equation}
\bm{k}\cdot\bm{l}=-1\quad{\rm and}\quad\bm{m}\cdot\bm{\overline{m}}=1.
\label{normc}
\end{equation}
   
The null vector $\bm{k}$ must be proportional to $\partial_v\bm{x}$~\cite{KSMH}. Taking into account the normalization conditions~(\ref{normc}) we obtain
\begin{equation}
k^\mu=(0,2,0,0)\quad;\quad k_\mu=(-1,0,0,0).
\label{k1}
\end{equation}

For the vector $\bm{m}$ we take
   
\begin{equation}
m^\mu=(0,\alpha,\beta,\gamma),
\label{m1}
\end{equation}
where $\alpha$, $\beta$, and $\gamma$ may be complex.
The conditions~(\ref{orthc}),~(\ref{nullc}) and~(\ref{normc}) determines $\alpha$, $\beta$ and $\gamma$. Thus,
\begin{equation}
\alpha=\frac{2}{\sqrt{2}}\left(\dot{\rho}+i\rho\dot{\phi}\right)\quad,\quad\beta=\frac{1}{\sqrt{2}}\quad,\quad\gamma=\frac{i}{\sqrt{2}\rho}. 
\label{mcomp}
\end{equation}

By following the standard notation, we can write the null tetrad basis $\{\bm{e}_{(a)}\}=\{\bm{l},\bm{k},\bm{m},\bm{\overline{m}}\}$,  
$
\bm{e}_{(1)}=l^\mu\frac{\partial}{\partial \xi^\mu}\;,\;
\bm{e}_{(2)}=k^\mu\frac{\partial}{\partial \xi^\mu}\;,\;
\bm{e}_{(3)}=m^\mu\frac{\partial}{\partial \xi^\mu}\;,\;
\bm{e}_{(4)}=\overline{m}^\mu\frac{\partial}{\partial \xi^\mu},
$
and the dual basis $\{\bm{e}_{(a)}\}$,
$
\bm{e}_{(1)}=l_\mu d\xi^\mu\;,\;
\bm{e}_{(2)}=k_\mu d\xi^\mu\;,\;
\bm{e}_{(3)}=m_\mu d\xi^\mu\;,\;
\bm{e}_{(4)}=\overline{m}_\mu d\xi^\mu,
$
where
\begin{subequations}
\label{l}
\begin{equation}
l^\mu=(\dot{u},\dot{v},\dot{\rho},\dot{\phi}),
\end{equation}
\begin{equation}
l_\mu=\left(\frac{1}{2}{\int \frac{\partial H}{\partial u}}du,-\frac{1}{2},\dot{\rho},\rho^2\dot{\phi}\right)
\end{equation}
\end{subequations}
\begin{subequations}
\label{k2}
\begin{equation}
k^\mu=(0,2,0,0),
\end{equation}
\begin{equation}
k_\mu =(-1,0,0,0),
\end{equation}
\end{subequations}

\begin{subequations}
\label{m}
\begin{equation}
m^\mu=\frac{1}{\sqrt{2}}\left(0,2(\dot{\rho}+i\rho\dot{\phi}),1,\frac{i}{\rho}\right),
\end{equation}
\begin{equation}
m_\mu=\frac{1}{\sqrt{2}}\left(-\frac{1}{2}(\dot{\rho}+i\rho\dot{\phi}),0,1,i\rho\right).
\end{equation}
\end{subequations}
Note that the components $l_\mu$ are the momenta canonically conjugated to the coordinates $(u,v,\rho,\phi)$.

Using the equations~(\ref{l})-(\ref{m}) and the definitions in the Appendix A, one can verify that the only non-necessarily null NP quantities are:
\begin{equation}
\Phi_{22}=\frac{1}{2}R_{(1)(1)}=-\frac{1}{4}\nabla^2_\perp H
\label{phi22}
\end{equation}
and
\begin{equation}
\Psi_4\!=\!C_{\mu\nu\gamma\delta}l^\mu \overline{m}^\nu l^\gamma \overline{m}^\delta  \!=\!\frac{1}{4}\!\left(\!\frac{H_{,\phi\phi}}{\rho^2}\!+\!\frac{H_{,\rho}}{\rho}-H_{,\rho\rho}\!\right)\!+\frac{i}{2}\!\left(\frac{H_{,\rho\phi}}{\rho}-\frac{H_{,\phi}}{\rho^2}\right)\!,
\label{psi4}
\end{equation}
where $C_{\mu\nu\gamma\delta}$ is the Weyl tensor and the partial derivatives are abbreviated by a comma. We can immediately see that if $H$ is a harmonic function of the transverse coordinates, $\Phi_{22}=0$. Moreover, if $H$ is cylindrically symmetric, $\Im m\Psi_4=0$.

\section{Geodesic deviation of test particles}
\label{secgd}
In this section we write the components of the Riemann tensor in terms of the NP quantities in a local orthogonal basis. We follow the approach of~\cite{BP1},  but we do not impose any field equation to the spacetime metric. Thus, we obtain a model independent description of geodesic deviation equation in $pp$ -wave spacetimes. 

Let ${\bm{u}}=d{\bf x}/d\tau$, $u_\mu u^{\mu}=-1$ be the four-velocity of free a test particle in a spacetime of curvature described by the Riemann tensor ${R^\mu }_{\nu\gamma\delta}$. The displacement vector $\bm{X}$ between two nearby particles must obey the geodesic deviation equation

\begin{equation}
\frac{D^2X^\mu }{d\tau^2}=-{R^\mu }_{\nu\gamma\delta}u^\nu  X^\gamma  u^\delta  ,
\label{geod1}
\end{equation}
where $\tau$ is the proper time of one of the particles.  

We define a local basis by the orthonormal tetrad  $\{\bm{e}_{\hat{a}}\}$, $\hat{a}=0,1,2,3$, where $\bm{e}_{\hat{0}}={\bm{u}}$ and $\{\bm{e}_{\hat{i}}\}$, $i=1,2,3$ are orthogonal space-like unit four-vectors, such that $\bm{e}_{\hat{a}}\cdot\bm{e}_{\hat{b}}\equiv g_{\mu\nu}e^\mu_{\hat{a}}e^\nu_{\hat{b}}=\eta_{\hat{a}\hat{b}}={\rm diag}(-1,1,1,1)$. The frame components are $X^{\hat{a}}=e^{\hat{a}}_\mu X^\mu$. The relative accelerations between two test particles in the local basis are defined by
\begin{equation}
\frac{d^2X^{\hat{a}}}{d\tau^2}=\bm{e}^{\hat{a}}\cdot\frac{D^2\bm{X}}{d\tau^2}.
\label{acc1}
\end{equation}
Then, we have
\begin{equation}
\frac{d^2X^{\hat{0}}}{d\tau^2}=R_{\mu\nu\gamma\delta}u^\mu  u^\nu  X^\gamma  u^\delta  =0
\label{zerocomp}
\end{equation}
and
\begin{equation}
\frac{d^2{X}^{\hat{i}}}{d\tau^2}=-{R^{\hat{i}}}_{\hat{0}\hat{j}\hat{0}}X^{\hat{j}},
\label{devinvar}
\end{equation}
where $X^{\hat{i}}\!\!=\!\!e^{\hat{i}}_\mu X^\mu $ gives the distance between two test particles and
\begin{equation}
R_{\hat{i}\hat{0}\hat{j}\hat{0}}~=~R_{\mu\nu\gamma\delta}e^\mu_{\hat{i}}e^\nu_{\hat{0}}e^\gamma_{\hat{j}}e^\delta_{\hat{0}}
\label{ProjRiem}
\end{equation}
are the projections of the Riemann tensor components on the local basis $\{\bm{e}_{\hat{a}}\}$.  
The local basis is related to a null tetrad basis $\{\bm{l},\bm{k},\bm{m},\bm{\overline{m}}\}$ by~\cite{KSMH}:
\begin{eqnarray}
{\bm{u}}=\frac{1}{\sqrt{2}}\left(\bm{l}+\bm{k}\right)\quad,\quad\bm{e}_{\hat{3}}=\frac{1}{\sqrt{2}}\left(\bm{k}-\bm{l}\right)\quad, \nonumber \\
\bm{e}_{\hat{1}}=\frac{1}{\sqrt{2}}\left(\bm{\overline{m}}+\bm{m}\right)\quad,\quad\bm{e}_{\hat{2}}=i\frac{1}{\sqrt{2}}\left(\bm{\overline{m}}-\bm{m}\right).
\label{basechang}
\end{eqnarray}
From the definition of the Weyl tensor we have
\begin{equation}
R_{\hat{i}\hat{0}\hat{j}\hat{0}}=C_{\hat{i}\hat{0}\hat{j}\hat{0}}+\frac{1}{2}(\delta_{ij}R_{\hat{0}\hat{0}}-R_{\hat{i}\hat{j}})-\frac{1}{6}\delta_{ij}R.
\label{Rdec}
\end{equation}
By using~(\ref{basechang}) and~(\ref{Rdec}), we can write the components of Riemann tensor on the local basis in terms of the twelve NP quantities. The result is in the Appendix A, equations~(\ref{Riemnp}). 
For the $pp$-wave spacetime the only non-necessarily null NP quantities are $\Phi_{22}$ and $\Psi_4$. Thus, the equations~(\ref{Riemnp}) reduces to:

\begin{subequations}
\label{Riemnppp}
\begin{equation}
R_{\hat{1}\hat{0}\hat{1}\hat{0}}=\frac{1}{2}\Re e\Psi_4+\frac{1}{2}\Phi_{22}\;,
\label{Riemnppp1}
\end{equation}
\begin{equation}
R_{\hat{1}\hat{0}\hat{2}\hat{0}}=-\frac{1}{2}\Im m\Psi_4\;,
\end{equation}
\begin{equation}
R_{\hat{2}\hat{0}\hat{2}\hat{0}}=-\frac{1}{2}\Re e\Psi_4+\frac{1}{2}\Phi_{22}\;.
\end{equation}
\end{subequations}
The equations~(\ref{devinvar}) reads
\begin{subequations}
\label{d2X}
\begin{equation}
\frac{d^2X^{\hat{1}}}{dt^2}=-(A_++A_\circ)X^{\hat{1}}+A_\times X^{\hat{2}},
\end{equation}
\begin{equation}
\frac{d^2X^{\hat{2}}}{dt^2}=A_\times X^{\hat{1}}-(-A_++A_\circ)X^{\hat{2}},
\end{equation}
\begin{equation}
\frac{d^2X^{\hat{3}}}{dt^2}=0,
\end{equation}
\end{subequations}
where $A_+\equiv\frac{1}{2}\Re e\Psi_4$, $A_\times\equiv\frac{1}{2}\Im m\Psi_4$ and $A_\circ\equiv\frac{1}{2}\Phi_{22}$. 

As can be immediately seen from~(\ref{d2X}), the generic $pp$-wave produces no variation in the longitudinal direction~$\bm{e}_{\hat{3}}$.
Under a rotation of the transverse plane by an angle $\vartheta$, $\bm{m}$ changes according to  $\bm{m}^\prime={\rm e}^{-i\vartheta}\bm{m}$. Then,
\begin{equation}
A_+^\prime=\cos(2\vartheta)A_+-\sin(2\vartheta)A_\times,
\end{equation}
\begin{equation}
A_\times^\prime=\cos(2\vartheta)A_\times+\sin(2\vartheta)A_+,
\end{equation}
\begin{equation}
A_\circ^\prime=A_\circ.
\label{ampvar}
\end{equation}
By using the above equations, it is easy to see that the equations~(\ref{d2X}) are invariant under rotations of $\vartheta=n\pi$, where $n$ is an integer. As $A_\circ$ is invariant under arbitrary rotations of the transverse plane we conclude that $A_+$ and $A_\times$ are responsible to the ``$+$'' and ``$\times$'' helicity-2 polarization modes and $A_\circ$ to the helicity-0 mode. By a rotation of $\vartheta=n\pi/4$ we have $A_+^\prime=-A_\times$ and $A_\times^\prime=A_+$. Then, a general observer sees a superposition of two helicity-2 polarization modes shifted by $\pi/4$ and one helicity-0 polarization mode. From~(\ref{phi22}) we can see that, if~$\nabla_\perp H=0$, there are no helicity-0 polarization modes.
As can be seen from~(\ref{psi4}), if the transverse space is cylindrically symmetric, $A_\times=0$ and the wave is purely ``$+$'' polarized.
Note that, we have not made any assumptions concerning the gravitational field equations or the underling theoretical model. The results obtained until  now follows directly from the structure of the $pp$-waves.

We also notice that, the line element~(\ref{lin1}) (or equivalently~(\ref{lin2})) is of  Kerr-Schild form~\cite{KSMH}. This means that, the spacetime metric $g_{\mu\nu}$ can be written as
\begin{equation}
g_{\mu\nu}=\eta_{\mu\nu}+Hp_\mu p_\nu,
\label{KSmetric}
\end{equation}
where  $p_\mu\equiv-\delta_{\mu u}$.
Then, the weak field regime is obtained simply by taken $H\ll1$. 
Note that since the $pp$-waves are exact solutions to general relativity and even to more general gravitational theories~\cite{buchdahl}, the weak field regime described here are also exact solutions since, $H$ are not imposed to satisfy the linearized gravity equations.

\section{Wave polarizations in quadratic curvature gravity}
\label{secqg}

In this section, we apply the formalism developed in the two previous sections to analyze the relative acceleration of test particles in presence of $pp$-waves in quadratic curvature gravity. To a review of the quadratic curvature gravity see for example the reference~\cite{schmidt} and the references cited there.

Consider the quadratic gravitational theory defined by the action
\begin{eqnarray}
S =\frac{1}{16\pi G}{\int d^4x\sqrt{-g}\{R+\alpha
R^2+\beta R_{\mu\nu}R^{\mu\nu}+16\pi G {\cal L}_m}\},\label{ac1}
\end{eqnarray}
where ${\cal
L}_m$ is the Lagrangian of matter fields and $G$ the Newton's gravitational constant. In writing the action~(\ref{ac1}) we make use of the four dimensional Gauss-Bonnet invariant to eliminate the quadratic invariant, $R_{\mu\nu\rho\sigma}R^{\mu\nu\rho\sigma}$ as is usual in the treatment of the quadratic curvature action in four dimensions~\cite{mannh}.

The field equations derived from $\delta S=0$ are,
\begin{eqnarray}
R_{\mu\nu}-\frac{1}{2}g_{\mu\nu}R+\alpha {\rm H}_{\mu\nu}+\beta
{\rm I}_{\mu\nu}=8\pi GT_{\mu\nu},\label{fieldeq}
\end{eqnarray}
where
${\rm H}_{\mu\nu}$ and ${\rm I}_{\mu\nu}$ are 
given by
\begin{equation}
{\rm H}_{\mu\nu}=-2 R_{;\mu\nu}+2g_{\mu\nu}\Box
R-\frac{1}{2}g_{\mu\nu}R^2+2RR_{\mu\nu},
\end{equation}
and
\begin{eqnarray}
{\rm I}_{\mu\nu}=-2R^\alpha_{\;\;\mu;\nu\alpha}+\Box
R_{\mu\nu}+\frac{1}{2}g_{\mu\nu}\Box
R+2R^\alpha_{\;\;\mu}R_{\alpha\nu}-
\frac{1}{2}g_{\mu\nu}R_{\alpha\beta}R^{\alpha\beta},
\end{eqnarray}
where $\Box$ is the curved space d'Alembert operator. To the theory defined by~(\ref{ac1}) to have an acceptable Newtonian limit, the parameters $\alpha$ and $\beta$ must satisfy the following constraints~\cite{PTP}:
\begin{equation}
3\alpha+\beta>0\quad,\quad\beta<0.
\label{abconstraints}
\end{equation}

For metrics of the form~(\ref{lin1}) we have $\Box\rightarrow\nabla_\perp^2$, where $\nabla_\perp^2$ is the Laplacian in the transverse space. Moreover, by virtue of~(\ref{0inv2}) there is no contribution of the ${\rm H}_{\mu\nu}$ term and the only contribution of the ${\rm I}_{\mu\nu}$ comes from $\Box
R_{\mu\nu}$. Therefore,
the field equations reduces to:
\begin{equation}
-\frac{1}{2}[\beta\nabla_\perp^4+\nabla_\perp^2]H(x_\perp,u)=8\pi G T_{uu}.
\label{qeq}
\end{equation}
This equation can be integrated to give
\begin{equation}
[\beta\nabla_\perp^2+1]H(x_\perp,u)=H_{1}(x_\perp,u)+ah(x_\perp,u),
\label{1intH}
\end{equation}
where $h(x_\perp,u)$ is a harmonic function of the transverse coordinates, $a$ an arbitrary constant and
\begin{equation}
\nabla_\perp^2H_{1}(x_\perp,u)=-16\pi GT_{uu}.
\label{Eeq}
\end{equation}
Now, we write $H(x_\perp,u)$ as
\begin{equation}
H(x_\perp,u)=H_{1}(x_\perp,u)+H_{2}(x_\perp,u)+ah(x_\perp,u),
\label{Hdec}
\end{equation}
where $H_{2}(x_\perp,u)$ is determined by the equation
\begin{equation}
\left[\nabla_\perp^2+\frac{1}{\beta}\right]H_{2}(x_\perp,u)=16\pi GT_{uu}.
\label{Hqeq}
\end{equation}
The functions $H_1$ and $H_2$ are respectively the purely linear and purely quadratic parts of the solution. The $pp$-wave solution to quadratic gravity is given by the metric~(\ref{lin1}) in which $H$ is given by the equation~(\ref{Hdec}).

Regarding the equation~(\ref{phi22}) and substituting~(\ref{Hdec}) in~(\ref{1intH}) we obtain
\begin{equation}
A_\circ\equiv\frac{1}{2}\Phi_{22}=-\frac{1}{8}\nabla_\perp^2H=\frac{1}{8\beta}H_{2}(x_\perp,u).
\label{Aoq}
\end{equation}
Thus, the helicity-0 component of the wave is given by the solutions of the equation~(\ref{Hqeq}) multiplied by $-1/\beta$, the inverse of the coupling parameter of the $R_{\mu\nu}R^{\mu\nu}$ invariant in the quadratic gravitational action. We notice that the linear curvature term do not contribute to the $\Phi_{22}$. Therefore, the transverse helicity-0 mode comes only from the quadratic curvature terms  and is determined by the equation~(\ref{Hqeq}). This fact contrasts with the result that $\Phi_{22}=0$ in empty space $pp$-wave solutions to Einstein's gravity. There are also also the nonvanishing components of helicity-2 namely, $A_+$ and $A_\times$ which can be computed from~(\ref{psi4}), where the effect of the quadratic curvature comes from the $H_2$ term in~(\ref{Hdec}). This result indicates that can be a contribution of quadratic curvature terms to the Einsteinian polarizations of a null wave. This fact provides a justificative, from a non perturbative point of view, to the appearance of a $\beta$ dependent correction in the amplitude of the linearized gravitational waves in the transverse traceless gauge when quadratic curvature terms are considered as small corrections to the Einstein's general relativity~\cite{RAA2}.

\subsection{An example of impulsive gravitational wave in quadratic gravity}

Let us give an explicit example of a solution obtained in quadratic gravity for a source consisting of a finite thin shell of width $2R_0$ and homogeneous energy density $\varrho_0$, which propagates with the light velocity in the $z$ direction. The $T_{uu}$ component of the matter energy-momentum tensor is: 

\begin{equation}
T_{uu}=\lambda\varrho_0\Theta(R_0-\rho)\delta(u),
\label{Tuu}
\end{equation}
where $\Theta(x)$ is the standard step function, $\rho$ is the radial coordinate from the source axis 
and $\lambda$ is a constant~\cite{RAA1}.
By solving the equations~(\ref{Eeq}) and~(\ref{Hqeq}) with $T_{uu}$ given by~(\ref{Tuu}) we obtain

\begin{eqnarray}
H(u,\rho)&=&\!\kappa\Bigg\{\!\left[4R_0bK_1\!\left(\frac{R_0}{b}\right)I_0\!\left(\frac{\rho}{b}\right)-\rho^2-4b^2\right]\!\Theta(R_0\!-\!\rho)\delta(u) \nonumber \\
&-&\!\!\!\!\left[2R_0^2\!\left(\!\ln\!\left(\frac{\rho}{R_0}\right)\!+\frac{1}{2}\right)\!+4R_0bI_1\!\left(\frac{R_0}{b}\right)\!K_0\!\left(\frac{\rho}{b}\right)\right]\!\Theta(\rho\!-\!R_0)\delta(u)\!\Bigg\},
\label{H(rho,u)}
\end{eqnarray}
where  $\kappa=4\pi G\lambda\varrho_0$, $K_\nu$ and $I_\nu$ are modified Bessel functions and $b\equiv\sqrt{-\beta}$ (see~\cite{RAA1} for the explanation of the boundary and regularity conditions which are imposed in the derivation of the result~(\ref{H(rho,u)})). The cylindrical symmetry of $H$ is due to the symmetry of the source~(\ref{Tuu}). We assume that $\beta<0$ since when $\beta>0$ there is no acceptable Newtonian limit for the nonrelativistic gravitational potential between point masses in quadratic gravity~\cite{PTP}. This choice excludes the non physical solutions which appear when $\beta>0$ leading to imaginary components in $H$.
Note that the solution $H$ is a continuous function of $\rho$ and diverges logarithmically for $\rho\rightarrow\infty$. Although $H$ is natural quantity to be taken as the amplitude of a $pp$-wave, this choice is not appropriate to the observational point of view since it contradicts the expectation that the  wave amplitude must decrease with the distance form the source. Thus, the quantities that can better represent wave amplitudes from the observational point of view must be given by the $A_+$, $A_\times$ and $A_0$ which determines the relative accelerations between the test particles. 

Using the equations~(\ref{phi22}) and~(\ref{psi4}) we obtain
\begin{eqnarray}
A_\circ=\frac{1}{2}\Phi_{22}&&=-\frac{\kappa}{2}\Bigg\{\left[\frac{R_0}{b}K_1\!\left(\frac{R_0}{b}\right)I_0\!\left(\frac{\rho}{b}\right)-1\right]\!\Theta(R_0\!-\!\rho)\delta(u)\nonumber \\
&&-\frac{R_0}{b}I_1\!\left(\frac{R_0}{b}\right)K_0\!\left(\frac{\rho}{b}\right)\!\Theta(\rho\!-\!R_0)\delta(u)\Bigg\},
\label{A_0}
\end{eqnarray}
\begin{eqnarray}
A_{+}\!=\!\frac{1}{2}\Re e\Psi_4\!&&=\!-\frac{1}{2}\kappa\Bigg\{\frac{R_0}{b}K_1\!\left(\frac{R_0}{b}\right)I_2\!\left(\frac{\rho}{b}\right)\!\Theta(R_0\!-\!\rho)\delta(u)\nonumber \\
&&+\left[\frac{R_0^2}{\rho^2}-\frac{R_0}{b}I_1\left(\frac{R_0}{b}\right)K_2\left(\frac{\rho}{b}\right)\!\right]\!\Theta(\rho\!-\!R_0)\delta(u)\Bigg\}
\label{A+}
\end{eqnarray}
and
\begin{equation}
A_\times=\frac{1}{2}\Im m\Psi_4=0.
\end{equation}
Both $A_\circ$ and $A_+$ are continuous functions of $\rho$ and goes to $0$ as $\rho\rightarrow\infty$. If an observer is placed at the source axis ($\rho=0$) with one of his frame direction, for instance the $z$ axis, aligned with the propagation direction of the wave, he do not see the helicity-2 component in the acceleration pattern of test particles since this observer is located at the symmetry axis of the source which has cylindrical symmetry. For this observer  $A_+=A_\times=0$ and the only nonvanishing pattern in the relative accelerations of test particles comes from
\begin{equation}
A_0(\rho=0)=-\frac{\kappa}{2}\left[\frac{R_0}{b}K_1\left(\frac{R_0}{b}\right)-1\right]\delta(u)=-\frac{1}{b^2}H(u,0)=-\frac{1}{b^2}H_2(0,u),
\label{H(0)}
\end{equation}
which has no contribution from the linear curvature (Einsteinian) part of the theory.
If the observer keep his frame orientation but, is displaced at a distance $\rho$ from the source axis, he observes only one component of helicity-2, given by~(\ref{A+}) in addition to the helicity-0 one given by~(\ref{A_0}).

\section{Summary and conclusions}

We have studied the deviation of geodesics in $pp$-wave spacetimes by relating the nonvanishing NP quantities of a general $pp$-wave in four spacetime dimensions with the Riemann tensor components in a local orthonormal basis. We have showed that the $pp$-wave solutions to quadratic curvature gravity produces relative accelerations between test particles located in the geodesics of the spacetime which are transverse to the wave propagation direction. These accelerations follow a pattern given by, at most, two components of helicity-2, analogous to the polarization pattern of a plane gravitational wave in linearized Einstein's gravity, and one of helicity-0. For a general quadratic gravity $pp$-wave we have obtained that there is a helicity-0 pattern in the relative accelerations of test particles which depends only on the purely quadratic part of the spacetime metric, namely $H_2$. A particular example of an impulsive $pp$-wave solution to quadratic gravity with cylindrical symmetry was given for which we identify one helicity-2 and one helicity-0 nonvanishing components that can be observed in the relative accelerations of nearby test particles. The suppression of one of the helicity-2 patterns occur due the cylindrical symmetry of the source. For an observer placed at $\rho=0$, the helicity-2 component vanishes due to the symmetry of the source. For this observer only the helicity-0 pattern, which depends only on the quadratic curvature part of the metric, survives.

The approach by which the relative accelerations of nearby test particles in a local orthonormal basis was obtained can be used to obtain the geodesic deviations in more general spacetime metrics. An interesting study that can be carried out within the context of quadratic curvature gravity concerns the geodesic deviations of test particles in presence of non-null wave-like solutions to quadratic gravity such as those obtained in~\cite{MR1}. However, we left the investigation of this subject to be carried out in another future work.

\begin{acknowledgements}
{I am greatful to Dr. Odylio D. Aguiar and Dr. J. C. N. de Araujo for the critical reading of the manuscript. I would like to thank the Brazilian agency FAPESP for financial support (grants 00/10374-5).} 
\end{acknowledgements}
\appendix

\section{}

We transcribe in this appendix the definitions of the NP quantities for a null tetrad basis $\{\bm{l},\bm{k},\bm{m},\bm{\overline{m}}\}$ and write the Riemann tensor components in the local (observer) basis $\{\bm{e}_{\hat{a}}\}$ in terms of the NP quantities.

In the NP notation, the null tetrad components of the traceless Ricci tensor ($S_{\mu\nu}\equiv R_{\mu\nu}-g_{\mu\nu}R/4$) can be written in terms of three real and three complex scalars according to the definitions~\cite{KSMH}:

\begin{eqnarray}
\Phi_{00}\equiv\frac{1}{2}S_{\mu\nu}k^\mu k^\nu =\overline{\Phi}_{00}\;,\;
\Phi_{01}\equiv\frac{1}{2}S_{\mu\nu}k^\mu m^\nu =\overline{\Phi}_{10}\quad,\nonumber\\
\Phi_{02}\equiv\frac{1}{2}S_{\mu\nu}m^\mu m^\nu =\overline{\Phi}_{20}\;,\;
\Phi_{11}\equiv\frac{1}{4}S_{\mu\nu}(k^\mu l^\nu +m^\mu \overline{m}^\nu )=\overline{\Phi}_{11}\quad,\nonumber\\
\Phi_{12}\equiv\frac{1}{2}S_{\mu\nu}l^\mu m^\nu =\overline{\Phi}_{21}\;,\;
\Phi_{22}\equiv\frac{1}{2}S_{\mu\nu}l^\mu l^\nu =\overline{\Phi}_{22}\quad.
\label{Phidef}
\end{eqnarray}
The Ricci scalar is denoted by
\begin{equation}
\Lambda\equiv\frac{1}{24}R=\frac{1}{12}(R_{(3)(4)}-R_{(1)(2)}),
\label{Lambdef}
\end{equation}
where $R_{(a)(b)}=R_{\mu\nu}e^\mu_{(a)}e^\nu_{(b)}$.
The Weyl tensor components in the null tetrad basis can be written in terms of the five complex scalars~\cite{KSMH}:

\begin{eqnarray}
\Psi_0=C_{\mu\nu\gamma\delta}k^\mu m^\nu k^\gamma m^\delta  \quad,\quad\Psi_1=C_{\mu\nu\gamma\delta}k^\mu l^\nu k^\gamma m^\delta  \quad,\nonumber\\
\Psi_2=\frac{1}{2}C_{\mu\nu\gamma\delta}k^\mu l^\nu (k^\gamma l^\delta  -m^\gamma \overline{m}^\delta  )\quad,\quad\Psi_3=C_{\mu\nu\gamma\delta}l^\mu k^\nu l^\gamma \overline{m}^\delta  \quad,\nonumber\\
\Psi_4=C_{\mu\nu\gamma\delta}l^\mu \overline{m}^\nu l^\gamma \overline{m}^\delta  \;.
\label{Psidef}
\end{eqnarray}

The Riemann tensor components in the local basis $\{\bm{e}_{\hat{a}}\}$ in terms of the NP quantities are:

\begin{subequations}
\label{Riemnp}
\begin{equation}
R_{\hat{1}\hat{0}\hat{1}\hat{0}}\!=\!\frac{1}{2}\Re e\Psi_0\!+\!\frac{1}{2}\Re e\Psi_4\!-\!\Re e\Psi_2+\frac{1}{2}\Phi_{22}+\frac{1}{2}\Phi_{00}-\Re e\Phi_{02}\!-\!2\Lambda\;,
\end{equation}
\begin{equation}
R_{\hat{1}\hat{0}\hat{2}\hat{0}}\!=\!\frac{1}{2}\Im m\Psi_0-\frac{1}{2}\Im m \Psi_4-\Im m \Phi_{02}\;,
\end{equation}
\begin{equation}
R_{\hat{1}\hat{0}\hat{3}\hat{0}}\!=\!-\Re e\Psi_1+\Re e\Psi_3-\Re e\Phi_{01}+\Re e\Phi_{12}\;,
\end{equation}
\begin{equation}
R_{\hat{2}\hat{0}\hat{2}\hat{0}}\!=\!-\frac{1}{2}\Re e\Psi_0\!-\!\frac{1}{2}\Re e\Psi_4\!-\!\Re e\Psi_2+\frac{1}{2}\Phi_{22}+\frac{1}{2}\Phi_{00}+\Re e\Phi_{02}\!-\!2\Lambda\;,
\end{equation}
\begin{equation}
R_{\hat{2}\hat{0}\hat{3}\hat{0}}\!=\!-\Im m\Psi_1-\Im m \Psi_3-\Im m \Phi_{01}+\Im m\Phi_{12}\;,
\end{equation}
\begin{equation}
R_{\hat{3}\hat{0}\hat{3}\hat{0}}=2\Re e\Psi_2+2\Phi_{11}-2\Lambda\;.
\end{equation}
\end{subequations}


\begin{thebibliography}{99}
\addcontentsline{toc}{chapter}{References}
\bibitem{RAA1}
E. C. de Rey Neto, J. C. N. de Araujo and O. D. Aguiar, Class. Quantum Grav. {\bf 20}, 1479 (2003).
\bibitem{RAA2}
E. C. de Rey Neto, O. D. Aguiar and J. C. N. de Araujo,  Class. Quantum Grav. {\bf 20}, 2025 (2003).
\bibitem{Chris1}
D. Christodoulou, Phys. Rev. Lett {\bf 67}, 1486 (1991).
\bibitem{CVV}
F. Canfora, G. Vilasi and P. Vitale,  Phys. Lett. {\bf 545B},  373 (2002); F. Canfora, G. Vilasi and P. Vitale, gr-qc/0212024;  F. Canfora and G. Vilasi, gr-qc/0301083.
\bibitem{KSMH}
D. Kramer, H. Stephani, E. Herlt, and M. MacCallum {\it Exact Solutions of
Einstein's Field
Equations} (Cambridge University Press, London, 1980).
\bibitem{buchdahl}
H. A. Buchdahl 1983 J. Phys. A: Math. Gen. {\bf 16}, 1441 (1983).
\bibitem{MR1}
M. D. Roberts, Int. J. Mod. Phys. {\bf 9}, 167 (1994). 
\bibitem{P1}
F. A. E. Piran F A E 1957 Phys. Rev. {\bf 105}, 1089 (1957).
\bibitem{Stein}
R. Steinbauer, J. Math. Phys. {\bf 39}, 2201 (1998); M. Kunzinger and R. Steinbauer,  J. Math. Phys. {\bf 40}, 1479 (1999).
\bibitem{kerner}
A. Balakin, J. W. van Holten, R. Kerner, Class. Quantum Grav. {\bf 17}, 5009 (2000); A. Balakin, R. Kerner and J. P. S. Lemos, Class. Quantum Grav. {\bf 18}, 2217 (2001); R. Kerner, J. W. van Holten, R. Colistete Jr, Class. Quantum Grav. {\bf 18}, 4725 (2001).
\bibitem{S1}
K. S. Stelle, Phys. Rev D {\bf 16}, 953 (1977).
\bibitem{HaySam2}
K. Hayashi and T. Samura, Mod. Phys. Lett. {\bf A11}, 1023 (1996).
\bibitem{BP1}
J. Bi\v c\' ak and J. Podolsk\'y,  J. Math. Phys. {\bf 40}, 4506 (1999).
\bibitem{schmidt}
H.-J. Schmidt, Astron. Nachr. {\bf 307}, 339 (1986); gr-qc/0106037.
\bibitem{mannh}
P. D. Mannheim, Gen. Relativ. Gravit. {\bf 22}, 289 (1990).
\bibitem{PTP}
A. Accioly, A. Mukai and E. C. de Rey Neto, Prog. Theor. Phys {\bf 104}, 103 (2000).
\end{thebibliography}
\end{document}